\documentclass[useAMS,usenatbib]{mn2e}
\usepackage{graphicx,rotate,url,mathptmx,lscape,times,amsmath,color,multirow}
\usepackage{aas_macros}
\font\manual=manfnt at 7pt \def\dbend{\hbox{\raise0.9ex\hbox{\manual\char127\hspace{0.6em}}}}

\providecommand{\e}[1]{\ensuremath{\times 10^{#1}}}

\newcounter{INTERNALionstage}

\def\gtsim{\mathrel{\hbox{\rlap{\hbox{\lower4pt\hbox{$\sim$}}}\hbox{$>$}}}}
\def\lesssim{\mathrel{\hbox{\rlap{\hbox{\lower4pt\hbox{$\sim$}}}\hbox{$<$}}}}

\def\ga{{\rm\thinspace gauss}}

\def\pcc{{\rm\thinspace cm^{-3}}}

\def\la{\mbox{{\rm L}$\alpha$}}

\DeclareMathAlphabet{\vib}{OML}{cmm}{m}{it}

\title[H, He-like recombination spectra I]{H, He-like recombination spectra I:  
$l$-changing collisions for hydrogen}

\author[F. Guzm\'an et al.]
       {\parbox[]{6.0in}
        { F. Guzm\'an$^1$, N.R. Badnell$^2$,  R.J.R. Williams$^3$, P. A. M. van Hoof$^4$, 
	M. Chatzikos$^1$ and G.J. Ferland$^1$. \\
        \footnotesize
        $^1$Department of Physics and Astronomy, University of Kentucky, Lexington, KY 40506, USA.\\
        $^2$Department of Physics, University of Strathclyde, Glasgow G4 0NG, UK.\\
        $^3$AWE plc, Aldermaston, Reading RG7 4PR, UK.\\
        $^4$Royal Observatory of Belgium, Ringlaan 3, 1180 Brussels, Belgium.}
}

\date{
      Received }

\pagerange{\pageref{firstpage}--\pageref{lastpage}}
\pubyear{}

\begin{document}

\maketitle

\label{firstpage}

\begin{abstract}

\noindent
Hydrogen and helium emission lines in nebulae form by radiative recombination. This is a simple process
which, in principle, can be described to very high precision.
Ratios of He~I and H~I emission lines can be used to measure the He$^+$/H$^+$ abundance ratio
to the same precision as the recombination rate coefficients.
This paper investigates the controversy over the correct theory to describe dipole $l$-changing collisions
($nl\rightarrow nl'=l\pm 1$) between energy-degenerate states within an $n$-shell.
The work of  \cite{PengellySeaton1964} has, for half-a-century, been considered the definitive
study which ``solved'' the problem.
Recent work by \cite{VOS2012} recommended the use of rate coefficients 
from a semi-classical approximation which are nearly an order of magnitude smaller than those of 
\cite{PengellySeaton1964}, with the result that significantly higher densities are needed for the $nl$  
populations to come into local thermodynamic equilibrium.

Here, we compare predicted H~I emissivities from the two works and find widespread differences,
of up to $\approx 10$\%. This far exceeds the 1\% precision required to obtain the primordial
He/H abundance ratio from observations so as to constrain Big Bang cosmologies.
We recommend using the rate coefficients of \cite{PengellySeaton1964} for 
$l$-changing collisions, to describe the H recombination spectrum, based-on their 
quantum mechanical representation of the long-range dipole interaction.
\end{abstract}


\section{Introduction}
\label{intro}

Measurement of the primordial abundances of the lightest elements 
is one of the three decisive tests of Big Bang cosmology
\citep{1993ppc..book.....P}.
\citet{1967ApJ...148....3W} showed how the lightest elements could be created during the 
first few minutes after the Big Bang,
and how their abundances vary with cosmological parameters.
The IAU meeting \emph{Light Elements in the Universe}
\citep{2010IAUS..268.....C}
summarizes the current observational situation for the light elements, and
\citet{2010IAUS..268..163F} discuss recent progress on measurements of
the He/H abundance ratio.

The predicted range in the He/H abundance ratio is not large, 
so it must be measured with a precision
approaching 1\%  if
a definitive cosmological test is to be made. 
The abundance ratio is measured in low-metallicity H~II regions by using H~I and He~I recombination lines.
Both H~I and He~I radiative effective recombination rate coefficients must be known to at least this 1\% precision.
We have focused on computing He~I recombination rate coefficients, with the spectral simulation code Cloudy (last described by
\citet{CloudyReview13}), in a series of papers
 \citep{Bauman2005,Porter2005,
 PorterFerlandMacAdam2007,
 Porter.R09Uncertainties-in-theoretical-HeI-emissivities:-HII-regions,
 Porter.R12Improved-He-I-emissivities-in-the-case-B-approximation,
Porter.R13-CaseB-erratum}.

\citet{Porter.R09Uncertainties-in-theoretical-HeI-emissivities:-HII-regions} carried-out 
Monte Carlo calculations to estimate the effect of various uncertainties in the atomic
data upon the accuracy of the effective recombination rate coefficients.
Rate coefficients for collisions were found to be the most uncertain quantity, in  general.
That study estimated the uncertainty in the rate coefficients for $l$-changing collisions, 
the topic of this paper, to be between 20\% and 30\%.

All of our previous work with Cloudy used the semi-classical (SC) theory of \cite{Vrinceanu2001} for $l$-changing collisions.
\citet{VOS2012} (hereafter VOS12) have recently introduced a simplified version (SSC) which is more
convenient for the generation of rate coefficients.
These (SC and SSC) rate coefficients are nearly an order of magnitude smaller than those derived by
\citet{PengellySeaton1964}, hereafter PS64, using impact parameter theory in the Bethe approximation
and long-considered to be the definitive work. (PS64 treats only dipole transitions, which dominate 
usually, but the work of \cite{Vrinceanu2001} is applicable for higher multipole transitions, such
as quadrupole.)
The SC theory was actually derived from their rigorous quantum mechanical theory (QM) \cite{Vrinceanu2001},
but VOS12 presented no QM rate coefficients.
\citet{2015MNRAS.446.1864S} noted that the QM probabilities in VOS12 appeared to agree well with
those of PS64, at large impact parameters, and so saw no reason to prefer SC over PS64.
How do QM and PS64 rate coefficients compare?
These (SC vs PS64) differences of  nearly 1 dex are far more extreme than the uncertainties
estimated by \citet{Porter.R09Uncertainties-in-theoretical-HeI-emissivities:-HII-regions}
in their analysis of the error propagation.
What effect will these differences have on the final effective recombination rates?
In addition, classical studies of H~I and H-like recombination spectra by 
\citet{Hummer1987} and \citet{Storey1995} made use of PS64.

This paper focuses on (dipole) $l$-changing collisions for hydrogen.
This is a simpler case than He since the $nl$-states within an $n$-shell have very nearly
the same energies, so the PS64 or VOS12
theories can be applied for all $l$.  This is unlike the case of He where the low-$l$ states 
are not energy degenerate and must be handled differently, introducing additional uncertainty.

\section{Atomic physics of l-changing collisions}

Several collisional processes affect how a recombined electron cascades to lower levels and produces
recombination emission lines.
Collisional ionization and its inverse, three-body recombination, are important for very high $n$-shells,
and act to bring those populations into local thermodynamic equilibrium (LTE) with the continuum. 
Electron $n$-changing collisions  affect populations of  levels at lower-$n$. Collisions out of 
metastable levels, such as  H~I 2s, become important at low densities.

This paper discusses the third type of collision, dipole $l$-changing collisions within an $n$-shell: $nl\rightarrow nl'=l\pm 1$.
These states have nearly the same energy so that ``Stark collisions'', caused by the electric field of
slow-moving massive particles such as protons or alpha particles, are the fastest.
These collisions change the intra-shell angular momentum of the orbiting electron and are very effective at 
large impact parameters.  Furthermore, if there is no energy difference between the states,
such as those of a fixed-$n$ in hydrogen, then the transition becomes more efficient 
as the speed of the charged  projectile decreases.

Our goal is to compare the results of \citet{PengellySeaton1964} and VOS12 theories and recommend
which to adopt for use in recombination calculations.
There are two main sources of disagreement between these theories and we discuss them in this section:

\begin{itemize}

\item The cross sections do not agree with each other. 
The PS64 cross sections are larger than the SC data of VOS12.  
However, the QM probabilities which VOS12 present do 
agree with those of PS64 at large impact parameters. 
This is discussed in section \ref{sec:cross sections}.

\item The QM treatments (PS64 and VOS12) require a cut-off to prevent the 
cross section integral, of the probability over
impact parameter ($b$), from diverging as $b \rightarrow \infty$. 
The SC theory of VOS12 does not require such a cut-off; intrinsically,
the probability falls to zero at a finite impact parameter but one that
is in general much smaller than the physically motivated QM one.
The QM cut-off is discussed in section \ref{sec:cut-off}.

\end{itemize}

\subsection{l-changing collisional data}
\label{sec:cross sections}

Here we use different {\it ab initio} approaches to compute the cross sections for $l$-changing collisions. 
For half-a-century the approach introduced by PS64 has been the only reference. 
PS64 use the Bethe approximation (\citealp{Bethe1930,Burgess1992}) in the Impact Parameter method 
and give a simple formula for the total rate coefficient, $q_{nl}$, for $nl\to nl^\prime$ summed over $l^\prime= l \pm 1$: 

\begin{align}
q_{nl}=&9.93\times10^{-6}\text{cm}^3\text{s}^{-1}\left(\frac{\mu}{m_e}\right)^{1/2} \nonumber \\
\times&\frac{D_{nl}}{T^{1/2}}\left[11.54+\log_{10}\left(\frac{Tm_e}{D_{nl}\mu}\right)+2\log_{10}\left(R_c\right)\right] \,,
\label{eq:PS64}
\end{align}

\noindent where $\mu$ is the reduced mass of the system, 
$m_e$ is the electron mass, $T$ is the temperature and $R_c$ is an effective cut-off radius. 

The cut-off, $R_c$, arises because the probability integral over impact parameter, which
gives rise to the cross section, diverges as the maximum impact parameter tends to infinity.
We discuss the choices, introduced by PS64, in section \ref{sec:cut-off}. 

A second  cut-off, at small impact parameter, was used also by PS64 in their derivation of 
\eqref{eq:PS64} to avoid divergence of the probability as the impact parameter tends to zero. 
This second cut-off, at $R_1$, is defined by $P(b=R_1)= 1/2$ and
then $P(b\le R_1)\equiv 1/2$. It does not normally play an important role.

The line factor in \eqref{eq:PS64}, $D_{nl}$, is given by

\begin{equation}
D_{nl}=\sum_{l^\prime=l\pm 1} \frac{D_{ji}}{\omega_l} = \left(\frac{Z_p}{Z_t}\right)^2 6n^2\left(n^2-l^2-l-1\right),
\label{eq:Dnllpm1}
\end{equation}

\noindent where $Z_p$ and $Z_t$ denote the charges of the projectile and target core respectively, $\omega_l=2l+1$ is the statistical weight of the initial state and

\begin{equation}
D_{ji}=\frac{8}{3}Z_p^2S_{ji}\,,
\end{equation}

\noindent where $i$ and $j$ 
are the initial and final state indexes, and $S_{ji}$ is the dipole line strength \citep{PengellySeaton1964}.
For partial rate coefficients, $q_{nl\to nl^\prime}$, we require explicitly

\begin{equation} 
D_{nl\to nl^\prime}=\left(\frac{Z_p}{Z_t}\right)^26n^2l_>\left(n^2-l_>^2\right)\,,
\label{eq:Dnl}
\end{equation}

\noindent where $l_> = \text{max}(l,l^\prime)$ and $l^\prime=l\pm 1$ still. 

VOS12 build upon the
QM and SC treatments  proposed by \cite{Vrinceanu2001,Vrinceanu2001b} 
where an exact solution was obtained by exploiting the symmetries of the Runge-Lenz operator. 
The QM probability for any multipole $l\to l^\prime$ intra-shell transition is given by VOS12:

\begin{align}
P^{QM}_{nl\to nl^\prime}=&\left(2l^\prime+1\right)\sum_{L=\left|l-l^\prime\right|}^{n-1}\left(2L+1\right)\left\{
\begin{matrix}
l & l^\prime & L\\
j & j  & j
\end{matrix}
\right\}^2 \nonumber \\
\times&\frac{\left(L!\right)^2\left(n-L-1\right)!}{\left(n+L\right)!} \nonumber\\
\times&\left(2\sin\chi\right)^{2L}\left[C_{n-L-1}^{(L+1)}\left(\cos\chi\right)\right]^2\,,
\label{eq:VOS12QM}
\end{align}

\noindent where $C_n^{(l)}$ is an ultraspherical polynomial, and $\{\ldots\}$ denotes a 6-j symbol 
\citep{Wigner1959}, where $j=(n-1)/2$. The rotation angle, $\chi$, is defined as:

\begin{equation}
\cos \chi = \frac{1+\alpha\cos\left(\pi\sqrt{1+\alpha^2}\right)}{1+\alpha^2}
\label{eq:coschi}
\end{equation}

\noindent for straight-line trajectories, where the scattering parameter, $\alpha$, is given by

\begin{equation}
\alpha=\frac{3Z_pn\hbar}{2m_evb}
\end{equation}

\noindent and $v$ is the projectile speed. 

The cross sections are obtained from the integration of the probability over the impact parameters:

\begin{equation} 
\sigma^{QM}_{nl\to nl^\prime}=2\pi\int_0^\infty{P^{QM}_{nl\to nl^\prime} bdb}\,.
\label{eq:sigmaQM}
\end{equation}

\noindent The integral \eqref{eq:sigmaQM} again diverges in the dipole case ($\left|l-l^\prime\right|=1$),
as  $P^{QM}_{nl\to nl^\prime}\propto b^{-2\left|l-l^\prime\right|}$,
and a cut-off must be applied. This point will be discussed in section \ref{sec:cut-off}.  

SC probabilities were obtained as well by \cite{Vrinceanu2001} by defining a classical analogue
of the Runge-Lenz quantal vector. The probability, as given in VOS12, is:

\begin{align}
P^{SC}_{nl\to nl^\prime}=\frac{2l^\prime}{\pi\hbar n^2\sin\chi}
\begin{cases}
0, &\text{if } \left|\sin\chi\right|<\left|\sin\left(\eta_1-\eta_2\right)\right| \\
\\
\frac{K(B/A)}{\sqrt{A}}, &\text{if } \left|\sin\chi\right|>\left|\sin\left(\eta_1+\eta_2\right)\right| \\
\\
\frac{K(A/B)}{\sqrt{B}}, &\text{if } \left|\sin\chi\right|<\left|\sin\left(\eta_1+\eta_2\right)\right| 
\end{cases}
\label{eq:VOS12SC6}
\end{align}

\noindent where $K$ is the complete elliptical integral, $A=\sin^2\chi-\sin^2\left(\eta_1-\eta_2\right)$, 
$B=\sin^2\left(\eta_1+\eta_2\right)-\sin^2\left(\eta_1-\eta_2\right)$, and $\cos\eta_1=l/n$ and $\cos\eta_2=l^\prime/n$. 
 Note that \eqref{eq:VOS12SC6} is zero for $l^\prime=0$. Instead, we obtain 
$P^{SC}_{nl\to n0} = (1/2l)P^{SC}_{n0->nl}$ from detailed balance. 
The same result is obtained on replacing the classical statistical weight $2l^\prime$ in \eqref{eq:VOS12SC6} by the quantum one $(2l'+1)$.
The probability from eq.\eqref{eq:VOS12SC6} agrees with further classical trajectory Monte Carlo calculations performed by VOS12. 
They further simplified it 
by approximating the scaled angular momentum 
$l/n$ as a continuous variable. 
When that is done then thermal averaging can be carried-out analytically and the rate coefficient obtained is:

\begin{align}
q^{SC}_{nl\to nl^\prime}=&1.294\times10^{-5}\text{cm}^3\text{s}^{-1}\left(\frac{\mu}{m_eT}\right)^{1/2}
\left(\frac{Z_p}{Z_t}\right)^2\nonumber\\
\times&\frac{n^2\left[n^2\left(l+l^\prime\right)-l_<^2\left(l+l^\prime+2\left|\Delta l\right|\right)\right]}
{\left(l+1/2\right)\left|\Delta l\right|^3}\,,
\label{eq:VOS12SC9}
\end{align}

\noindent where $l_<=\text{min}(l,l^\prime)$. We note that VOS12 replaced the original factor $1/l$
that arises in the denominator of \eqref{eq:VOS12SC9} by $l+1/2$ so that the method can be applied to s-states.

We will refer to equation \eqref{eq:PS64} results as PS64, 
\eqref{eq:VOS12QM} results as VOS12-QM, \eqref{eq:VOS12SC6} as VOS12-SC and 
\eqref{eq:VOS12SC9} as VOS12-SSC, since they correspond to the QM, 
and the two semi-classical (SC and SSC) methods of the cited reference, respectively. 
The equations \eqref{eq:VOS12QM}, \eqref{eq:VOS12SC6} and \eqref{eq:VOS12SC9} are summarized in VOS12,
where the authors compare the SC values of equation \eqref{eq:VOS12SC9} with the previous results 
of PS64 corresponding to equation \eqref{eq:PS64}. They find that the latter are typically a 
factor 6 larger than the SC results and state that PS64 overestimates the values 
due to the use of the Born approximation. 

\begin{figure}
\begin{center}
\includegraphics[width=0.4\textwidth,clip]{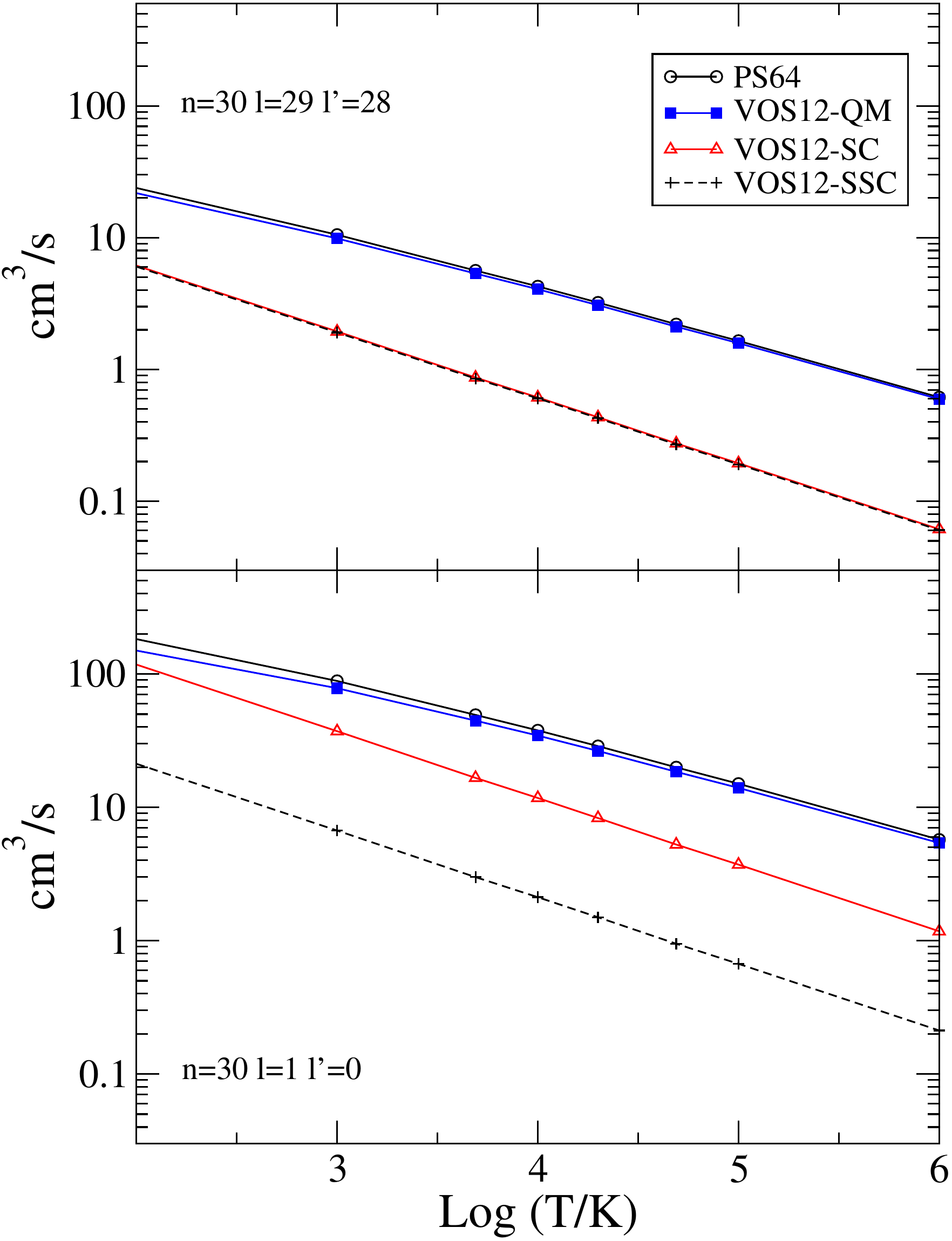}
\caption{\label{f:rates} Comparison of $l$-changing collisional rate coefficients for H$^+$+H$(n=30)$ 
collisions at high- and low-$l$. 
$n_\text{H}=10^4\text{cm}^{-3}$.
}
\end{center}
\end{figure}

We compare rate coefficients for the above methods  in Fig. \ref{f:rates} for $n=30$ and two 
$l\to l^\prime=l-1$ transitions: $l=29\to 28$ and $l=1\to 0$.
As in VOS12, the PS64 results are larger than the SC ones by a factor 5-10 from low to high temperatures for 
high $l$ and an order of magnitude larger than VOS12-SSC and a factor 2-5 larger
 than VOS12-SC for low $l$ from low to high temperatures. However, the QM results (blue symbols), of VOS12-QM, agree much better with PS64, differences ranging from 9-3\% for the high-$l$ case, and 18-6\% for the low-$l$, from 
low to high temperatures, respectively. The differences between SC and QM rate coefficients can be
explained by the contribution to the underlying probabilities from  large impact parameters. 

\begin{figure}
\begin{center}
\includegraphics[width=0.4\textwidth,clip]{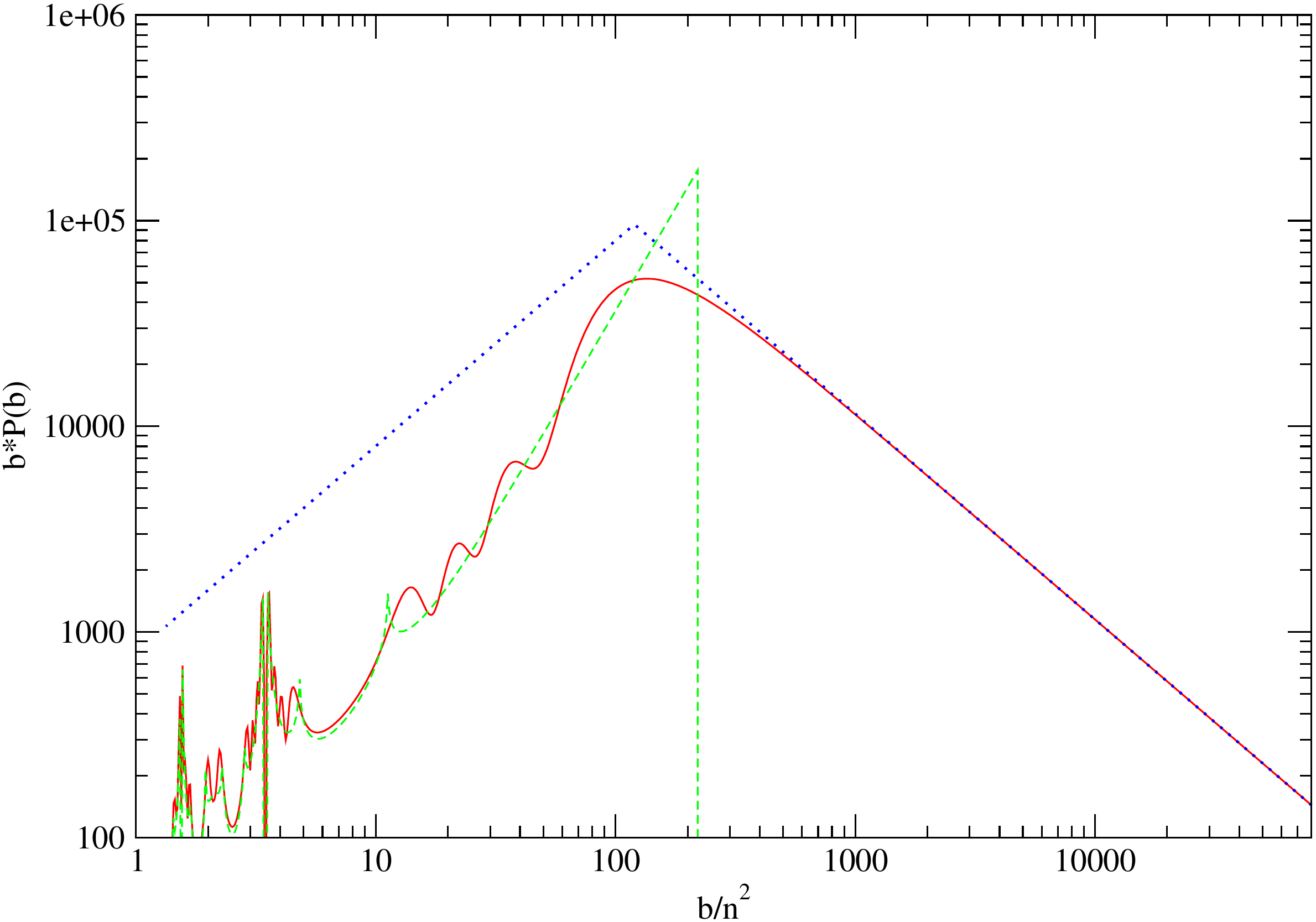}
\caption{\label{f:pvsb} Probability times the impact parameter versus scaled impact parameter 
for $n=40$ $l=36\to35$, at $v=0.25/n$. Solid line, VOS12-QM; dashed line, VOS12-SC; dotted line, PS64.}
\end{center}
\end{figure}

In Fig. \ref{f:pvsb}, we plot the impact parameter times probability,
since that is the relevant quantity for cross sections, for $n=40$ $l=36\to35$. 
The maximum impact parameter corresponds to a Debye radius (i.e. $R_c$) for $T/N_e=10^{-2}$~K.cm$^3$.
We see that the SC probabilities fall abruptly to zero at a finite impact-parameter, given by \eqref{eq:VOS12SC6}, 
which can be interpreted as an intrinsic  cut-off. In contrast, the QM probabilities fall-off smoothly, accounting
for the  long-range dipole interaction. While VOS12 did not compare rate coefficients from the SC and QM methods,
they did compare probability integrals (effectively, cross sections) for $n=40$  and several $l$-values and a
range of multipoles, and apparently very good agreement can be seen, in VOS12 fig. 2.
However, close inspection of their figure reveals that there are no QM results for dipole transitions: 
$\Delta l/n = \pm 0.025$.

The large difference between SC and QM dipole rates can be viewed as a failure 
of SC to fully account for the contribution from larger impact parameters to the long-range dipole interaction.
We see no reason to question the quantum mechanical treatment here and the ultimate lifetime or Debye cut-off
has a sound physical basis (which we discuss next). In this conclusion, we concur with \citet{2015MNRAS.446.1864S}.

\subsection{QM probability cut-off}
\label{sec:cut-off}

The electrostatic dipole interaction is asymptotic to $1/r^2$. As such, distant encounters
play an important role in the collision process, whether they be described by an integral
over impact parameters or a sum over angular momenta in a partial wave expansion.
Indeed, both are mathematically (logarithmically) divergent. 
This theoretical logarithmic behaviour of the dipole cross section has been verified 
by comparison with experimental results, for Ca$^+$ and Ba$^+$ --- see \citet{Burgess1974}.
However, the collision
does not take place in splendid isolation. Nature eliminates the contribution from the
most distant encounters, thus rendering a finite contribution. To model this physical
situation a cut-off is introduced at large impact parameters, both to the upper integral limit
in equation \eqref{eq:sigmaQM} and in the corresponding integral in PS64 which leads to equation \eqref{eq:PS64}.

PS64 propose two types of cut-off in the case of degenerate $l$-states. 
Firstly, the initial level must not radiate before the collision is complete. 
Using a fixed lifetime $\tau$ for the initial level, the maximum impact parameter would be:

\begin{equation}
R_L  = 0.72 v\tau\,,
\label{eq:Rc1}
\end{equation}

\noindent where $v$ is the projectile speed still. 

The second possible cut-off comes from the collective screening effect of the other 
electronic charges in the plasma and is equal to the Debye radius\footnote{The Debye radius 
is underestimated, by a factor $\sqrt{\pi}$, in equation (33) of PS64.}:

\begin{equation}
R_D = \left[\frac{kT_e}{4\pi e^2N_e}\right]^{1/2}\,.
\label{eq:RD}
\end{equation}

The smaller of these two cut-offs gives the maximum impact parameter which can contribute to the collision. 
These cut-offs affect both QM formalisms, PS64 and VOS12-QM. 
When the cut-off is large enough, the differences at small impact parameters (see Fig.  \ref{f:pvsb}) have a negligible
effect since large impact parameters dominate and
are responsible for the disagreement of the VOS12-QM with SC results, as shown in figures \ref{f:rates} and \ref{f:pvsb}. 
As Debye screening depends on density, rate coefficients may differ appreciably 
at high densities, compared to low, since the Debye radius is much smaller then.
 Tests on the sensitivity of the rate coefficients to the precise value of the 
cut-off typically gave differences of $\sim$10\% ($\sim$5\% for high $l$) in the rate 
coefficients for a variation of 20\% in the cut-off, at the highest densities considered in this work 
($n_\text{H}=10^{10}\text{cm}^{-3}$). At densities relevant to where the data have an
influence on the emissivities, $n_\text{H}=10^4\text{cm}^{-3}$ (see Fig. 
\ref{f:linesratio}), the differences in the rate coefficients are $\lesssim$2\%.

\section{Cloudy simulations}

 In order to clarify the impact of the $l$-changing  data on H~I emission lines 
 we have performed a series of tests that compared the line emission 
 resulting from using rate coefficients from the four cases represented by equations \eqref{eq:PS64}, \eqref{eq:VOS12QM}, 
\eqref{eq:VOS12SC6} and \eqref{eq:VOS12SC9}. 
For equation \eqref{eq:VOS12QM} we  applied the PS64 cut-off criteria to avoid divergence again at large impact parameters. 
For both \eqref{eq:VOS12QM} and \eqref{eq:VOS12SC6} averaging over the Maxwell distribution has been performed numerically. 

\subsection{Description of the model for H}

We use the development version of the spectral simulation code Cloudy C13 (branch Hlike\_HS87 revision r11051, experimental). 
We focus on ``Case B'' \citep{1938ApJ....88...52B}, in which it is assumed that higher-$n$ Lyman
lines scatter often enough to be degraded into a Balmer line plus Ly$\alpha$. 
This gives a surprisingly accurate description of H~I emission in low-density nebulae
\citep{AGN3}.
We compute hydrogen-only clouds, where Lyman lines are given a large optical depth.
The cloud is irradiated by a narrow band of light (a ``laser'') at 1.1Ry to provide ionization but without
pumping the H~I lines.  An ionization parameter \citep{AGN3} of $U = 0.1$ is assumed. 
The gas kinetic temperature of 1\e{4}~K is assumed. 
Various hydrogen densities are assumed and the
electron density is  taken to be equal to the hydrogen density for these conditions. 

Rate coefficients for electron impact excitation of $n\leq 5, l<n$ states are taken from 
\cite{Anderson2000,Anderson2002}.
Other excitation, ionization and three-body recombination rate coefficients are taken from \cite{Vriens1980}. 
Data for excitation by proton impact ($n$-changing) collisions are taken as well from \citet{Vriens1980}. 
This set of data are ``standard'' in Cloudy simulations and have been assessed elsewhere \citep{Porter2005,Bauman2005}.

Electronic configurations up to $n=200$ are included in the simulations. 
For the greatest accuracy and optimal computational speed, 
we assume that populations of very highly-excited levels, 
where collisional $l$-changing  transitions are faster than radiative decays, 
are statistically populated within $n$ (PS64). 
For a given $n$, the critical density $N_\text{crit}(n)$ for this to happen is:

\begin{equation}
N_\text{crit}(n)=\left(q_{n}\tau_{n}\right)^{-1}\,,
\label{eq:ncrit}
\end{equation}

\noindent where $q_{n}$ is the collisional rate $q_{nl}$
averaged over $l$, and $\tau_{n}$ is the $l$-averaged radiative lifetime. 
Levels above this are ``collapsed" while those below it are ``resolved'', as described by
\citet{CloudyReview13}.
Collapsed levels are treated as unresolved for $n$-changing transitions. 

Densities range between $10 \pcc$ to $10^{10} \pcc$ covering diverse astrophysical
applications from the interstellar medium to quasars \citep{AGN3}. 
At low densities, the maximum principal quantum number of resolved levels, $n_r$, is $\sim 60$, 
while at higher densities we keep $n_r\ge 20$ to ensure high accuracy.

Solutions for the level populations and resulting emission are obtained. 
A set of 107 H~I lines were selected to cover the wavelength range from 2000\AA\ 
to 30$\mu$m.  
This covers much of the spectral range where today's observatories can detect H~I emission. 

\subsection{Results}

\begin{figure*}
\begin{center}
\includegraphics[width=0.5\textwidth,clip]{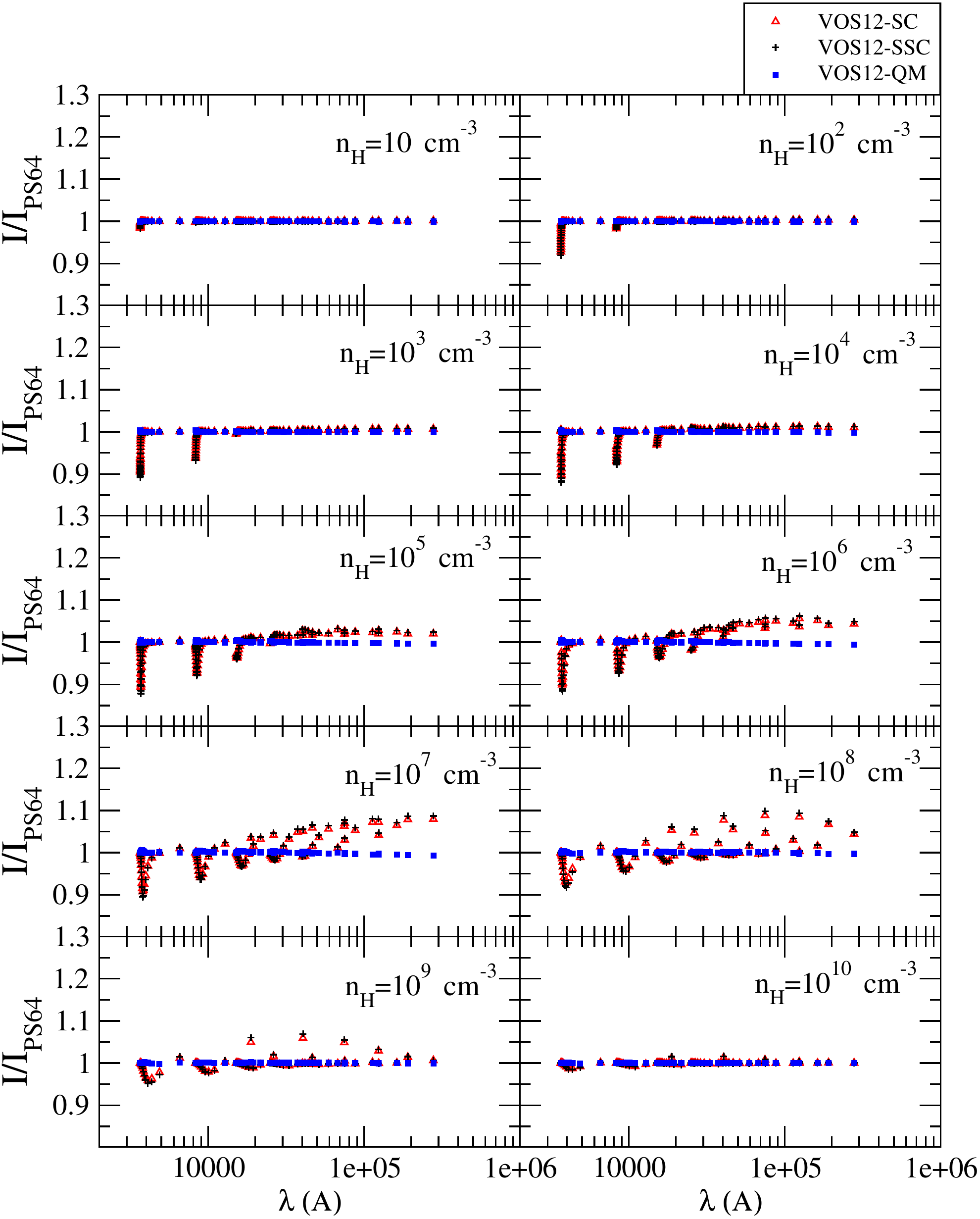}
\caption{\label{f:linesratio} Ratios of H~I lines with respect to the PS64 approach,
at $T=10^4$K, for the Cloudy simulations obtained using the different 
$l$-changing collisional data sets considered in this work.}
\end{center}
\end{figure*}

The effect of the variation of $l$-changing rates on the intensities of emission lines can be seen in Fig. \ref{f:linesratio}.
We show the predicted emission for each theory, relative to the emission predicted assuming the PS64 data.
The Figure shows a range of densities, which increases from left to right and from the top to the bottom.
The ratios are shown as a function of wavelength to make it easy to identify the various
converging Balmer, Paschen, etc. series. 

Emissivities are very similar for low densities, where the transitions are mostly radiative, 
and for high densities, where collisional LTE
has been reached. In these asymptotic cases differences in $l$-changing collision  have little effect 
on the final populations and emission results.

At intermediate densities, competition between collisional and radiative processes 
cause $l$-changing collisions to have greater influence on some lines. 
Differences of up to 10\%, much larger than the accuracy we seek,
are found when switching between the SC and QM treatments. 
Decay to $n\geq 2$ series, starting with the Balmer series, are clearly visible 
as the lower $n$-shells  have only a few  $l$-subshells available. 
Dipole decays to these levels will be from $l\pm 1$,
so the effects of $l$-changing collisions upon higher level populations affects these lines. 
In other words, the higher the $n$ of the lower level, the more $l$ subshells become available for 
redistribution and differences on the $l$-changing collision data become more important. 
As Fig. \ref{f:rates} suggests, results based in QM calculations agree more closely 
with each other than for SC calculations, where there are larger deviations. 

Balmer and Paschen lines are weaker in the SC case. 
This is probably caused by the smaller redistribution from higher-$l$ populations at high-$n$
due to the smaller SC $l$-changing dipole rates, as low-$l$ subshells are the only ones that can radiate
via a dipole transition to Balmer or Paschen states ($n=2$ and $n=3$).  
For larger values of the lower-$n$ of the transition, there is  a higher number 
of $l$-subshells available. The more effective redistribution of the QM 
treatments results in a larger number of decay routes becoming available, which in turn causes high-$l$ 
decay lines to be weaker than in the SC case.
Our results give differences (for QM vs SC) up to a 10\% for all the lines considered here.          

We also tested  the influence of various assumptions for the $l$-changing collision data upon
the intensities of the Lyman series.
Case A calculations, in which all lines are optically thin, were also performed. 
Deviations were small and mainly occur at high densities for the Ly$\alpha$,
where up to $\sim8$\% differences are obtained for the VOS12-SSC set of data 
for $n_\text{H} \sim 10^6\text{cm}^{-3}$. Other lines changed at the 0-1\% level. 
The 2s$\to 1$s two-photon transition has a much lower probability of decay than the direct 2p--1s. So,
it is clear that deviations in $l$-changing rate coefficients within the $n=2$ shell could have a stronger 
influence on Ly$\alpha$ line intensity than on other higher lines in the Lyman series. 
That happens at higher densities because of the high critical density for $n=2$.

\section{Conclusions}

We have compared the H~I emission spectra resulting from the  use of four sets of
dipole $l$-changing collision data corresponding to the four theories now available,
two quantum mechanical and two semiclassical. 
The two quantum mechanical treatments produce similar spectra, as do the two semiclassical treatments.
The Cloudy simulations show that the emissivities differ by less than 1\% 
between the two quantum mechanical theories.
However, we found that the $\sim 1$~dex differences between SC and QM treatments propagate up to 
a $\approx$ 10\% difference in the H~I emissivities. 
Which do we recommend?

The SC theory has the advantage that it is not sensitive to assumed cut-offs,
because it has an intrinsic, but unphysical, cut-off.
The QM theories require a cut-off because the probabilities diverge at large impact parameters.
This QM divergence has a physical basis due to the long-range dipole electrostatic interaction.
Based on the widely accepted divergent nature of dipole $l$-changing collisions, 
we see no reason to discard the quantum mechanical treatment.
In this, we concur with \citet{2015MNRAS.446.1864S}. 

After careful analysis of the quantal treatments, 
we consider that the cut-off criteria given by PS64, 
applied to the exact QM probabilities of VOS12-QM calculations, 
have a solid physical basis and yield accurate rate coefficients.
These are close to those of the Bethe approximation used by PS64. 
Thus, the results of \citet{Hummer1987} and \citet{Storey1995}, which made use of PS64,
are not affected by our findings.
The PS64 rates, being an analytic solution, are much faster to compute than the 
integration (in impact parameter and Maxwell distribution, as implied by $\alpha$) required for the 
VOS12 quantum mechanical treatment. 
Thus, we will use PS64 as the default for dipole $l$-changing rate coefficients and provide and option
to use the VOS12 quantum mechanical treatment when higher accuracy is required. 

The accuracy of the QM $l$-changing rate coefficients
used by Cloudy is now estimated to be well within the 20\% assumed by 
\citet{Porter.R09Uncertainties-in-theoretical-HeI-emissivities:-HII-regions} and so the resultant
H~I emissivities are accurate to the 1\% required if they are to be used to constrain Big Bang Cosmology models.

\section{Acknowledgments}

We thank D. Vrinceanu and H. Sadeghpour for helpful responses to several queries about their published work.
This work has been supported by the 
NSF (1108928, 1109061, and 1412155), NASA (10-ATP10-0053, 10-ADAP10-0073, NNX12AH73G, and ATP13-0153), and STScI
(HST-AR- 13245, GO-12560, HST-GO-12309, GO-13310.002-A, and HST-AR-13914). MC has been supported by STScI (HST-AR-14286.001-A).

\bibliographystyle{mn2e}
\bibliography{LocalBibliography,bibliography2.bib}
\bsp

\label{lastpage}
\clearpage
\end{document}